\documentclass[prc,twocolumn,showpacs,floatfix,nofootinbib,preprintnumbers,superscriptaddress,amsmath,amssymb]{revtex4-1}

\usepackage{CJKutf8}
\usepackage{graphicx}
\usepackage{epsfig}
\usepackage{bm}
\usepackage{color}
\usepackage{float}
\usepackage{dcolumn}
\usepackage{multirow}

\newcommand{\UNEDFONE}{\textsc{unedf1}}
\newcommand{\UNEDFONESO}{\textsc{unedf1}$^{\rm SO}$}
\newcommand{\HFODD}{\textsc{hfodd}}
\newcommand{\Nosc}{N_{\rm osc}}

\begin{document}

\begin{CJK*}{UTF8}{gbsn}

\title{Structure of Superheavy Nuclei Along Element 115 Decay Chains}

\author{Yue Shi (石跃)}
\affiliation{Department of Physics and Astronomy, University of Tennessee, Knoxville, Tennessee 37996, USA}
\affiliation{Joint Institute for Heavy Ion Research, Oak Ridge National Laboratory, Oak Ridge, Tennessee 37831, USA}

\author{D.E. Ward}
\affiliation{Division of Mathematical Physics, LTH, Lund University, SE-22100, Lund, Sweden}

\author{B.G. Carlsson}
\affiliation{Division of Mathematical Physics, LTH, Lund University, SE-22100, Lund, Sweden}

\author{J. Dobaczewski}
\affiliation{Institute of Theoretical Physics, Faculty of Physics, University of Warsaw, ul. Ho{\.z}a 69, PL-00681 Warsaw, Poland}
\affiliation{Department of Physics, PO Box 35 (YFL), FI-40014
University of Jyv{\"a}skyl{\"a}, Finland}

\author{W. Nazarewicz}
\affiliation{Department of Physics and Astronomy, University of
Tennessee, Knoxville, Tennessee 37996, USA}
\affiliation{Physics Division, Oak Ridge National Laboratory, Oak Ridge, Tennessee 37831, USA}
\affiliation{Institute of Theoretical Physics, Faculty of Physics, University of Warsaw, ul. Ho{\.z}a 69, PL-00681 Warsaw, Poland}

\author{I. Ragnarsson}
\affiliation{Division of Mathematical Physics, LTH, Lund University,  SE-22100, Lund, Sweden}

\author{D. Rudolph}
\affiliation{Department of Physics, Lund University, SE-22100, Lund, Sweden}

\begin{abstract}

A recent  high-resolution $\alpha$, $X$-ray, and $\gamma$-ray
coincidence-spectroscopy experiment offered a first glimpse  of excitation
schemes of isotopes along $\alpha$-decay chains of $Z=115$. To understand these
observations and to make predictions about shell structure of superheavy
nuclei below $^{288}115$, we employ two complementary mean-field models:
self-consistent Skyrme Energy Density Functional approach and the
macroscopic-microscopic Nilsson model. We discuss the spectroscopic
information carried by the new data. In particular, candidates for the
experimentally observed $E1$ transitions in $^{276}$Mt are  proposed. We find
that the presence and nature of low-energy $E1$ transitions in well-deformed
nuclei around $Z=110, N=168$ strongly depends on the strength of the
spin-orbit coupling; hence, it provides an excellent constraint on theoretical
models of superheavy nuclei. To clarify competing theoretical scenarios, an
experimental  search for $E1$ transitions in  odd-$A$ systems $^{275,277}$Mt,
$^{275}$Hs, and $^{277}$Ds is strongly recommended.

\end{abstract}


\maketitle

\end{CJK*}

\section{Introduction}

Superheavy nuclei at the limit of nuclear mass and atomic number pose a
formidable challenge to both experiment and theory. The low cross sections for
production of these nuclei, in the picobarn range or less, offer limited
structural information. Moreover, the $\alpha$-decay chains of nuclei
synthesized in experiments using a $^{48}$Ca beam with actinide targets
\cite{(Oga07),(Oga11),(Oga12),(Oga12a),(Oga13),(Dul10),(Gat11),(Hof12),(Rud13a)}
terminate by spontaneous fission before reaching the known region of the
nuclear chart. This poses a problem with the unambiguous identification of the
new isotopes, and more direct techniques to determine $Z$ and $A$ must be
employed \cite{(Rud13a)}. Theoretical predictions of the shell structure of
superheavy nuclei are also difficult, as the interplay between the
electrostatic repulsion and nuclear attraction, combined with a very high
density of single-particle (s.p.) states, make the results of calculations
extremely sensitive to model details
\cite{(Cwi96),(Kru00),(Ben01),(Cwi05),(Ben13),(Shi13a)}.

In a recent experimental study \cite{(Rud13a),(Rud13b)}, unique structural
information on low-lying states in superheavy nuclei below $^{288}115$ has been
obtained. Of particular interest is the finding that some of the measured
transitions in the nucleus assigned to be $^{276}$Mt have $E1$ character, thus
suggesting opposite parities of the connected states. The new data offer an
exciting opportunity to constraint theoretical models in this region for the
first time. Indeed, previous macroscopic-microscopic
\cite{(Cwi94),(Par04),(Par05)} and self-consistent studies
\cite{(Cwi99),(Ben00b)} have shown that the number of opposite-parity
s.p.~orbitals around the Fermi level is fairly limited, and this is consistent
with the Nilsson model analysis of Ref.~\cite{(Rud13a)}.

Because of the above-mentioned sensitivity to model details, robust predictions
in this region are difficult to make as one is dealing with large
extrapolations. To this end, when aiming at reliable predictions, it is
advisable to use a model that performs well in the neighboring region where
experimental information is more abundant. Furthermore, since the quadrupole
deformations of $\alpha$-decay daughters of $^{288}115$ are expected to
increase gradually with decreasing $Z$ and $A$ along the $\alpha$-decay chain
\cite{(Cwi94),(Par04),(Par05),(Cwi99),(Ben00b),(Cwi05),(Ben13)}, shape
polarization is going to play a role when determining the energies of low-lying
states.

In this work, we study the low-lying states in the superheavy nuclei below
$^{288}115$, using the locally-optimized self-consistent Skyrme Energy Density
Functional (SEDF) and Nilsson-Strutinsky (NS) frameworks. To assess the
robustness of these results, we also carry out calculations using a
globally-optimized SEDF model.

\section{Models}

The SEDF approach is a variant of nuclear density functional theory, which
offers a global, self-consistent description of nuclear properties across the
nuclear landscape \cite{(Ben03),(Erl12)}. The recent self-consistent study of
Ref.~\cite{(Shi13a)} offers a locally optimized SEDF parameterization
{\UNEDFONESO} that meets our local-extrapolability requirements: it reproduces
one-quasiparticle (1-q.p.) states in $^{251}$Cf and $^{249}$Bk (the two
heaviest systems where 1-q.p.~energies are experimentally well known), predicts
crucial deformed shell gaps at $N = 152$ and $Z = 100$, and describes
rotational bands in Fm, No, and Rf isotopes. The parameter set {\UNEDFONESO}
has been obtained by adjusting the spin-orbit coupling constants of a global
SEDF parametrization {\UNEDFONE} \cite{(Kor12)} that performs well for heavy
nuclei and large deformations. We shall also use {\UNEDFONE} in this study. The
calculations follow closely Ref.~\cite{(Shi13a)}. The Skyrme
Hartree-Fock-Bogolyubov (SHFB) equations were solved using the
symmetry-unrestricted solver {\HFODD} (v2.52j) \cite{(Sch12)} by expanding
1-q.p. wave functions in 680 deformed harmonic-oscillator (HO) basis states. To
compute 1-q.p.~excitations in odd-$A$ nuclei, we blocked relevant orbits around
the Fermi level as described in Ref.~\cite{(Sch10)}. The strengths of the
pairing force for neutrons and protons were adjusted to the odd-even mass
staggering in $^{251}$Cf and $^{249}$Bk and the kinematic moment of inertia of
$^{252}$No.

The SEDF results are compared with those of the Nilsson-Strutinsky (NS)
approach of Ref.~\cite{(Car06)} with the modified harmonic oscillator (MO)
potential and pairing as in Ref.~\cite{(Nil69)}. The shell-independent MO
parameters ($\kappa_p = 0.058$, $\mu_p = 0.63$, $\kappa_n = 0.0526$, and $\mu_n
= 0.457$) have been locally optimized to the actinide nuclei \cite{(Roz86)} and
applied  to, e.g., $^{228,230}$Pa \cite{(Her89)} and $^{242}$Am \cite{(Hay10)}.

The 2-q.p.-plus-rotor calculations for odd-odd nuclei were carried out using
the MO model of Ref.~\cite{(Rag88)}. The moments of inertia were chosen
according to a phenomenological relation of Ref.~\cite{(Gro62)}. The BCS
pairing was treated as in Ref.~\cite{(Nil69)}, with the monopole pairing
strengths taken as 95\% of the values for even-even nuclei. No residual
proton-neutron interaction was considered.

\section{Results}

\begin{figure}[htb]
\includegraphics[width=1.0\columnwidth]{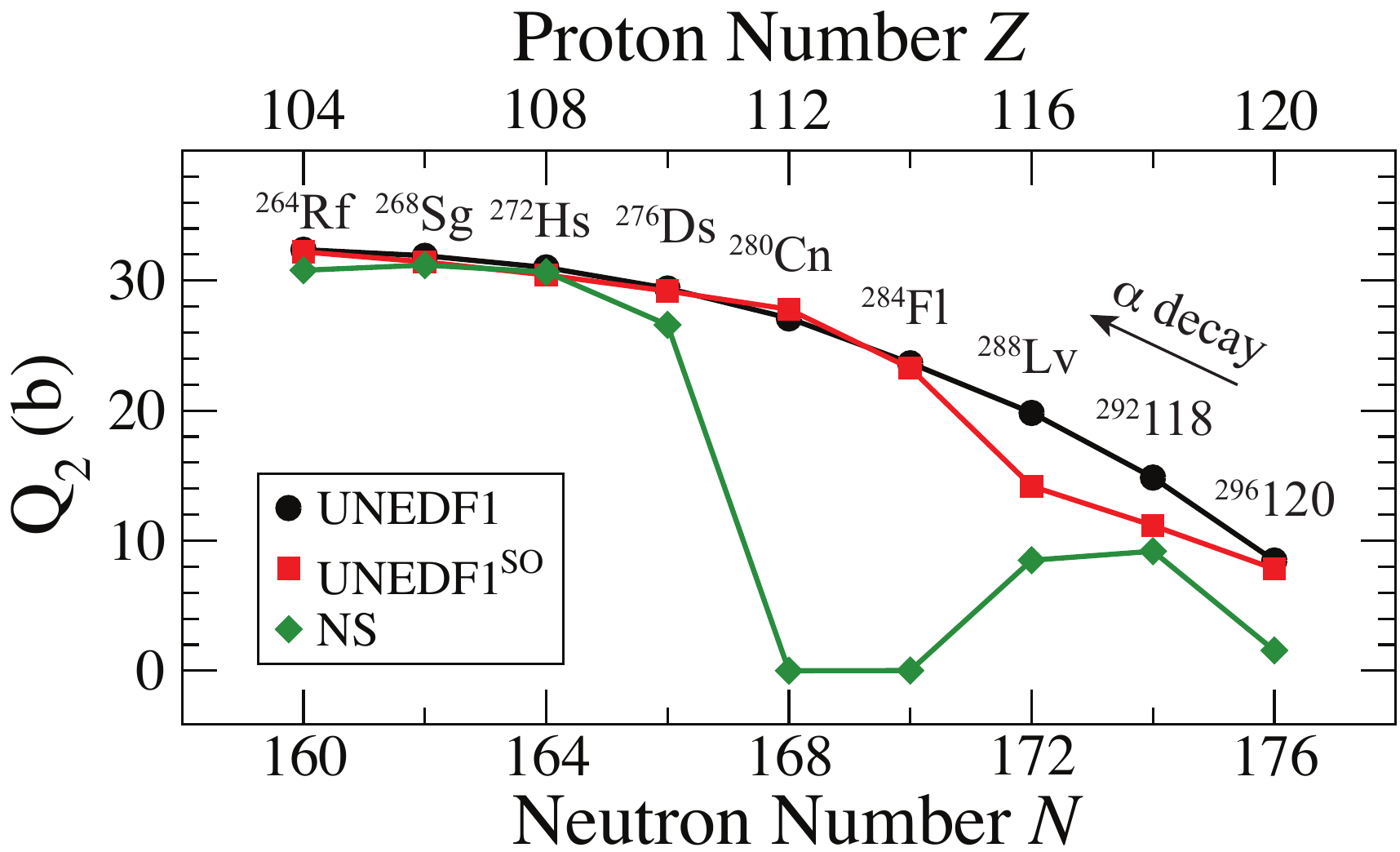}
\caption{\label{figQ2} (Color online) Quadrupole moments $Q_2$ of even-even
nuclei forming the $\alpha$-decay chain $^{296}120\rightarrow \cdots
\rightarrow ^{264}$Rf, calculated with {\UNEDFONESO} and {\UNEDFONE} SEDF models
and the NS approach.
}
\end{figure}

We first discuss properties of the even-even nuclei belonging to the
$\alpha$-decay chain of $^{296}$120. Their ground states form q.p.~vacua for
neighboring odd-$A$ and odd-odd systems. The calculated quadrupole moments are
shown in Fig.~\ref{figQ2}. Both SEDF models predict a similar smooth increase
of quadrupole deformation along the $\alpha$-chain. In the NS calculations,
$^{296}$120 is nearly spherical, $^{292}$118 and $^{288}$Lv are very weakly
deformed, $^{284}$Fl and $^{280}$Cn are spherical, and the shapes of the
lightest daughters have deformations close to those predicted by SEDF. These
results suggest that a direct comparison between SEDF and NS models is most
meaningful for $Z < 112$.

\begin{figure}[htb]
\includegraphics[width=0.8\columnwidth]{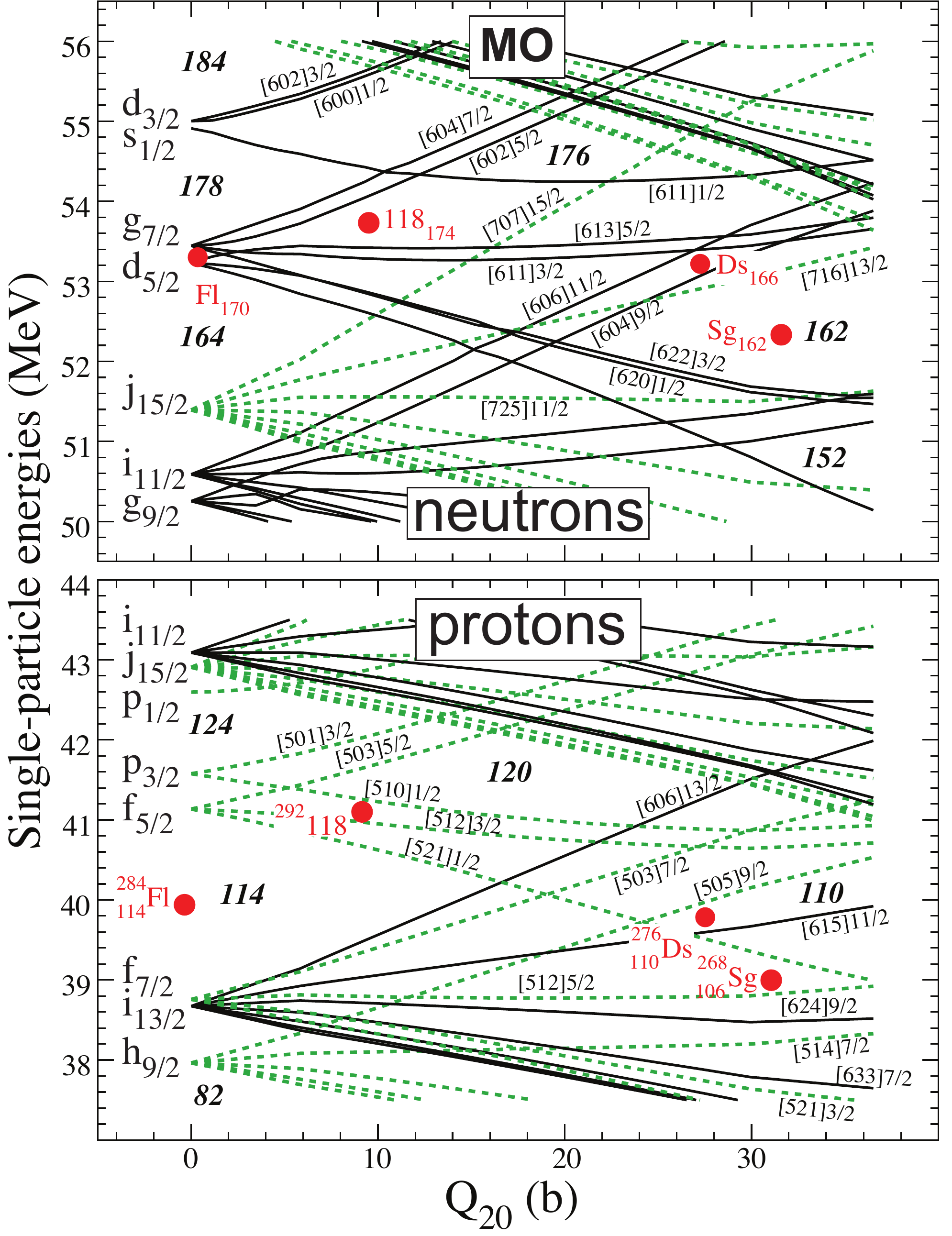}
\caption{
\label{figEspNS} (Color online) Nilsson diagram for neutrons
(top) and protons (bottom) for nuclei along the $\alpha$-decay chain of
$^{296}$120 using the MO potential of Ref.~\cite{(Roz86)}. The orbits are
labelled by the standard asymptotic Nilsson numbers \cite{(Nil69)}. The positive/negative
parity levels are marked by solid/dashed lines. The Fermi levels of nuclei in
Fig.~\ref{figQ2} are indicated by dots. The quadrupole moment was determined
from shape deformations $\epsilon_2$ and $\epsilon_4$ \cite{defors}: $Q_{20}=0.8 AR_0^2
(\epsilon_2 + 0.5 \epsilon^2_2+0.758\epsilon^2_4-\epsilon_2 \epsilon_4)$ with
$A=280$, $R_0 = r_0 A^{1/3}$, and $r_0= 1.217$\,fm.
}
\end{figure}

It is instructive to begin the discussion from the Nilsson s.p.~diagram of the
MO potential shown in Fig.~\ref{figEspNS}. The main features of this diagram,
such as the appearance of spherical shell gaps at $Z=114$ and $N=184$, have
remained unchanged since the late 1960s \cite{(Gus67),(Mos69)}. It is worth
noting that the s.p. spectrum of the MO model, with its pronounced spherical
shell gaps at $Z=114$ and $N=184$ and resulting Nilsson orbits, is fairly close
to that of more realistic Woods-Saxon \cite{(Cwi94),(Par04),(Par05)} and
Folded-Yukawa \cite{(Bol72),(Nix72),(Mol94)} potentials, see
Refs.~\cite{(Cwi96),(Ben99a)} for more discussion.

The deformed shell structure of nuclei at the end of the $\alpha$-decay chain
of $^{296}$120 (or $^{288}115$) is relatively simple: both in neutrons and
protons there appears one unique-parity, high-$\Omega$ Nilsson state
($\nu$[716]13/2 and $\pi$[615]11/2) surrounded by levels of opposite parity,
such as neutron ([613]5/2, [611]3/2), ([606]11/2, [604]9/2) and proton
([503]7/2, [505]9/2), ([510]1/2, [512]3/2) pseudo-spin doublets, respectively.

The spherical shell structure in superheavy nuclei strongly depends on the
spin-orbit splitting, which governs the size of the  $Z=114$ gap (cf. Table 4
of Ref.~\cite{(Cwi96)} and discussion therein). Also, the coupling between
Coulomb interaction and nuclear interaction is expected to impact the
predictions. To consider both effects, we studied s.p.~canonical states
obtained with {\UNEDFONESO} and {\UNEDFONE} SEDF models, which  differ in the
spin-orbit sector and treat the electrostatic energy self-consistently.

\begin{figure}[htb]
\includegraphics[width=\columnwidth]{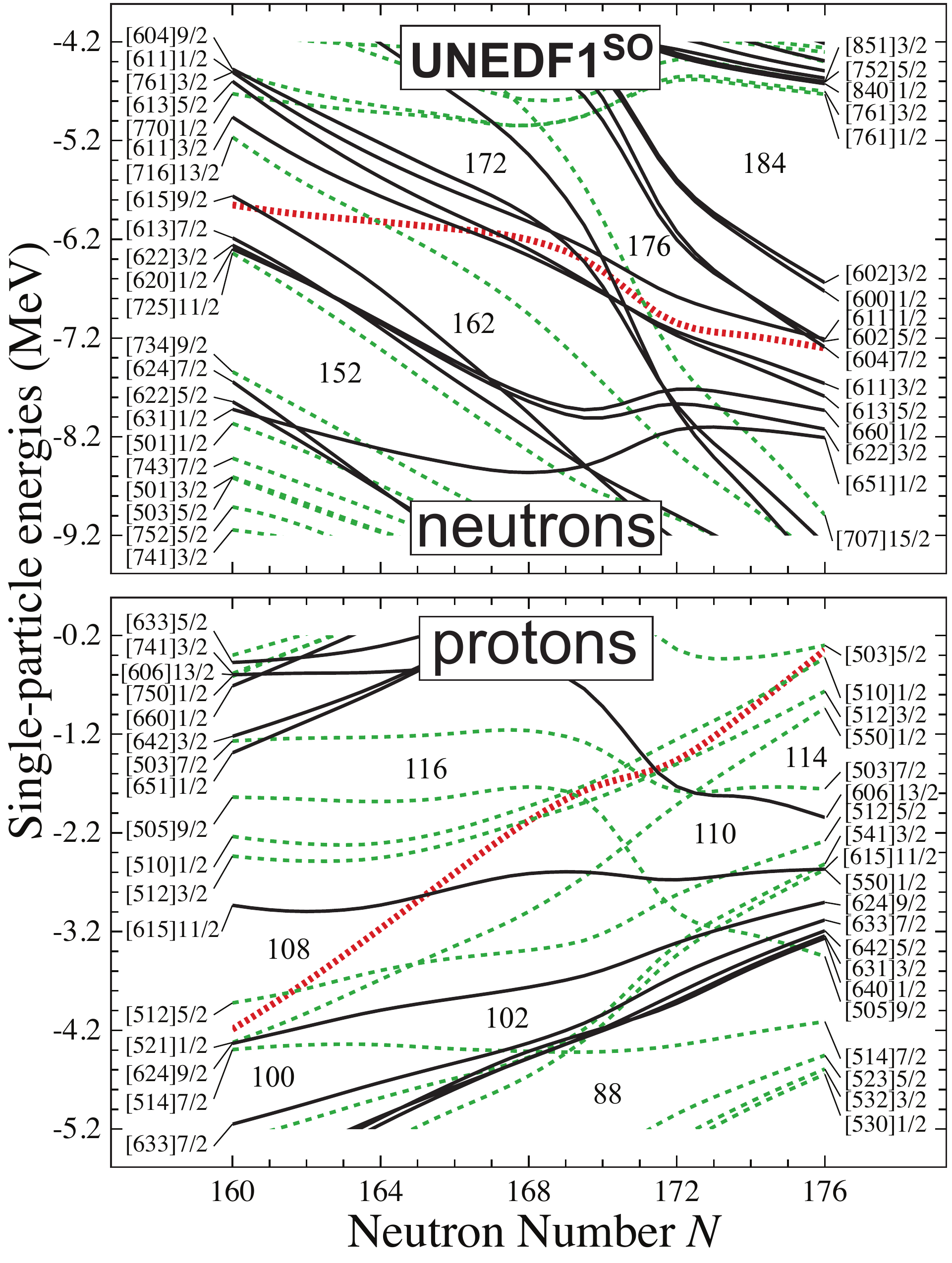}
\caption{\label{figEspEDF}
(Color online) Single-neutron (top) and single-proton (bottom) canonical
energies of {\UNEDFONESO} for nuclei along the $\alpha$-decay chain of
$^{296}$120 as in Fig.~\ref{figQ2}. The orbits are labelled by the standard
asymptotic Nilsson numbers corresponding to the dominant components of the SHFB
canonical wave functions. The positive/negative parity levels are marked by
solid/dashed lines. The Fermi levels are indicated by thick dotted lines.}
\end{figure}

The s.p.~energies of {\UNEDFONESO} along  the $\alpha$-decay chain of
$^{296}120$ are depicted in Fig.~\ref{figEspEDF}.

The s.p.~neutron spectrum is dominated by deformed gaps at $N=152$ and 162, and
a large spherical shell gap at $N=184$. In the deformed region $160 \le N \le
168$, the Nilsson states close to the Fermi level are primarily  $\Nosc=6$
levels and one unique-parity, high-$\Omega$ intruder level $\nu$[716]13/2
originating from the spherical 1$j_{15/2}$ shell. The structure of the proton
Nilsson diagram in Fig.~\ref{figEspEDF} is dominated by deformed gaps at
$Z=100$, 102, and 108, and a spherical subshell closure at $Z=114$. The
unique-parity, high-$\Omega$ intruder level $\pi$[615]11/2 originating from the
spherical 1$i_{13/2}$ shell is surrounded by several $\Nosc=5$ Nilsson
orbitals.

The general pattern of s.p.~states predicted by  {\UNEDFONESO} is not that far
from that in Fig.~\ref{figEspNS} of the  MO potential. However, there are
differences in the spherical shell structure, which will impact detailed
predictions for deformed superheavy nuclei belonging to $Z=115$ $\alpha$-decay
chains. In particular, MO predicts larger spherical shell gaps at  $Z=114$,
$N=148$, and $N=178$. In {\UNEDFONESO}, the splitting between the 1$j_{15/2}$
and 1$i_{11/2}$ spherical neutron shells is very small. This results in an
upward shift of the ([606]11/2, [604]9/2) doublet.

\begin{figure}[htb]
\includegraphics[width=\columnwidth]{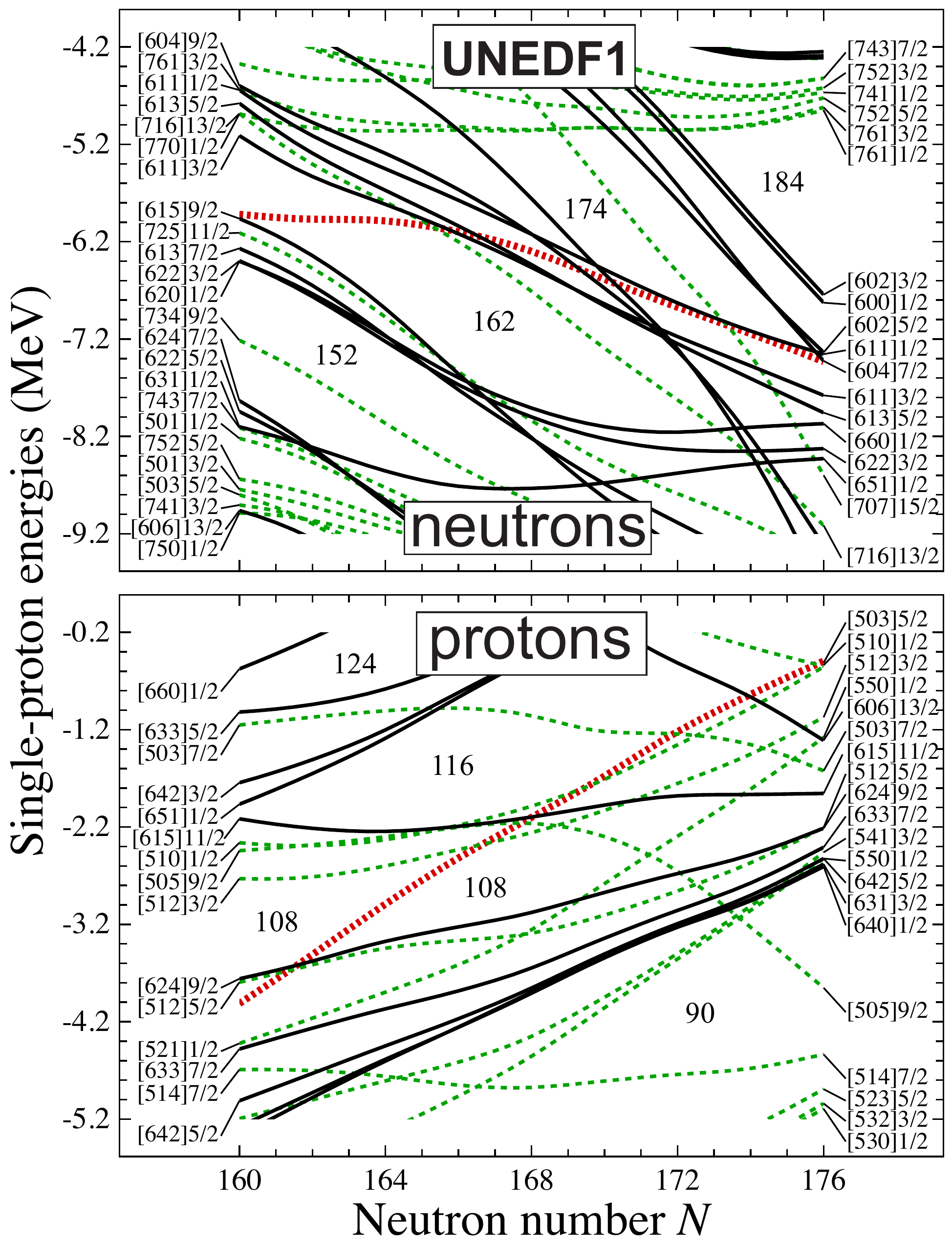}
\caption{\label{figEspEDForg}
(Color online) Similar as in Fig.~\ref{figEspEDF} but for {\UNEDFONE}}
\end{figure}

As seen in Fig.~\ref{figEspEDForg}, in the case of  {\UNEDFONE} the
unique-parity $\nu$1$j_{15/2}$ and $\pi$1$i_{13/2}$ shells are shifted up by a
few hundred keV, which results in a significant reduction of spherical $N=164$
and $Z=114$ shell closures \cite{(Shi13a)}. The change in the spin-orbit
potential also impacts positions of deformed levels. In particular, the
deformed neutron gap at $N=152$ is reduced, and that at $N=162$ opens up. In
the proton sector, the deformed Nilsson state [615]11/2 appears just below the
significantly increased $Z=116$ gap, close to the [505]9/2 and [510]1/2 levels.
The second proton intruder state [624]9/2 shows up  just below the deformed
proton gap at $Z=108$.

\subsection{One-quasi-particle energies}
To get more insights, we computed the energies of 1-q.p.~excitations for
odd-$Z$, even-$N$ superheavy nuclei that form the $\alpha$-decay chains of
$^{287}_{116}$Lv$_{171}$ and $^{289}_{116}$Lv$_{173}$ (Tables~\ref{tabN} and
\ref{tabN-orig}), and $^{287}115_{172}$ and $^{293}$117$_{176}$
(Tables~\ref{tabZ} and \ref{tabZ-ori}); the theoretical error on
1-q.p.~excitations due to the adopted size of the HO basis is less than 60\,keV
when going from 680 stretched HO states to 969 states. The results for
$^{287}_{116}$Lv$_{171}$ and $^{293}$117$_{176}$ are shown in Figs.~\ref{figN}
and \ref{figZ}, respectively.
\begin{table}[H]
\caption{\label{tabN}
Excitation energies, total quadrupole moments, and quadrupole mass
deformations $\beta_2$ for one-quasi-particle  excitations  in selected nuclei belonging to the
$\alpha$-decay chains of $^{287,289}$Lv predicted with  {\UNEDFONESO}. The
high-$\Omega$ unique-parity states are printed in boldface. The intrinsic
configurations are labelled as in Fig.~\ref{figEspEDF}. The predictions on the
lightest member of $^{289}$Lv $\alpha$-decay chain, $^{277}$Ds, are discussed below. 
}
\begin{ruledtabular}
\begin{tabular}{clccc}
Nucleus & Config.  &  $E_x$ (MeV) & $Q_{20}$\,(b) & $\beta_2$ \\
\hline  \\[-8pt]
$^{275}_{110}$Ds$_{165}$ & [611]3/2 & 0 & 29.8 & 0.23\\
                         & [613]5/2 & 0.085 & 29.7 & 0.23\\
                         & \textbf{[716]13/2} & 0.151 & 29.9 & 0.23 \\
                         & [611]1/2 & 0.305 & 29.8 & 0.23\\
                         & [604]9/2 & 0.619 & 29.4 & 0.22 \\
\hline  \\[-8pt]
$^{279}_{112}$Cn$_{167}$ & [611]3/2 & 0 & 28.3 & 0.21\\
                         & [613]5/2 & 0.013 & 28.3 & 0.21\\
                         & [611]1/2 & 0.121 & 28.3 & 0.21\\
                         & [604]9/2 & 0.306 & 28.0 & 0.21\\
                         & \textbf{[716]13/2} & 0.627 & 28.5 & 0.22\\
\hline  \\[-8pt]
$^{281}_{112}$Cn$_{169}$ & [611]1/2 & 0 & 27.2 & 0.20\\
                         & [604]9/2 & 0.145 & 27.4 & 0.21\\
                         & [613]5/2 & 0.159 & 27.1 & 0.20\\
                         & [611]3/2 & 0.237 & 26.2 & 0.20\\
                         & [606]11/2 & 0.606 & 24.8 & 0.19\\
\hline  \\[-8pt]
$^{283}_{114}$Fl$_{169}$ & [611]1/2 & 0 & 25.4 & 0.19\\
                         & [604]9/2 & 0.165 & 25.0 & 0.19\\
                         & [613]5/2 & 0.184 & 24.9 & 0.19\\
                         & [611]3/2 & 0.208 & 24.5 & 0.18\\
                         & [606]11/2 & 0.399 & 23.6 & 0.18\\
\hline  \\[-8pt]
$^{285}_{114}$Fl$_{171}$ & [611]1/2 & 0 & 21.0 & 0.15\\
                         & [613]5/2 & 0.051 & 19.0 & 0.14\\
                         & [611]3/2 & 0.056 & 19.0 & 0.14\\
                         & [606]11/2 & 0.058 & 21.0 & 0.15\\
                         & \textbf{[707]15/2} & 0.232 & 19.6 & 0.14\\
\hline  \\[-8pt]
$^{287}_{116}$Lv$_{171}$ & [613]5/2 & 0 & 14.2 & 0.10\\
                         & [611]3/2 & 0.085 & 14.2 & 0.10\\
                         & \textbf{[707]15/2} & 0.219 & 14.8 & 0.10\\
                         & [611]1/2 & 0.314 & 14.3 & 0.10\\
                         & [622]3/2 & 0.378 & 13.1 & 0.09\\
                         & [604]9/2 & 0.495 & 16.2 & 0.12\\
\hline  \\[-8pt]
$^{289}_{116}$Lv$_{173}$ & [613]5/2 & 0 & 11.8 & 0.08\\
                         & [611]3/2 & 0.029 & 11.8 & 0.08\\
                         & [611]1/2 & 0.055 & 11.9 & 0.08\\
                         & [604]7/2 & 0.372 & 11.2 & 0.08\\
                         & [602]5/2 & 0.397 & 11.3 & 0.08
\end{tabular}
\end{ruledtabular}
\end{table}

Although s.p.~energies are not experimental observables, those around the Fermi
level carry information about the low-lying q.p.~configurations in neighboring
odd-$A$ and odd-odd nuclei.

\begin{table}[H]
\caption{\label{tabN-orig} Similar as in Table~\ref{tabN} but for {\UNEDFONE}.
}
\begin{ruledtabular}
\begin{tabular}{clccc}
Nucleus & Config.  &  $E_x$ (MeV) & $Q_{20}$\,(b) & $\beta_2$ \\
\hline \\[-8pt]
$^{275}_{110}$Ds$_{165}$ & [613]5/2 & 0 & 30.2 & 0.23\\
                         & [611]3/2 & 0.050 & 30.2 & 0.23\\
                         & \textbf{[716]13/2} & 0.121 & 30.2 & 0.23 \\
                         & [611]1/2 & 0.364 & 30.2 & 0.23\\
                         & [604]9/2 & 0.561 & 29.8 & 0.23 \\
\hline \\[-8pt]
$^{279}_{112}$Cn$_{167}$ & [611]3/2 & 0 & 28.1 & 0.21\\
                         & [613]5/2 & 0.044 & 28.1 & 0.21\\
                         & [611]1/2 & 0.074 & 28.2 & 0.21\\
                         & \textbf{[716]13/2} & 0.197 & 28.3 & 0.21\\
                         & [604]9/2 & 0.201 & 27.8 & 0.21\\
\hline \\[-8pt]
$^{281}_{112}$Cn$_{169}$ & [611]1/2 & 0 & 26.3 & 0.19\\
                         & [604]9/2 & 0.109 & 26.2 & 0.19\\
                         & [611]3/2 & 0.152 & 25.7 & 0.19\\
                         & [613]5/2 & 0.157 & 26.0 & 0.19\\
                         & [606]11/2 & 0.267 & 25.0 & 0.19\\
\hline \\[-8pt]
$^{283}_{114}$Fl$_{169}$ & [611]1/2 & 0 & 25.5 & 0.19\\
                         & [604]9/2 & 0.114 & 25.1 & 0.19\\
                         & [611]3/2 & 0.147 & 24.7 & 0.18\\
                         & [613]5/2 & 0.160 & 24.6 & 0.18\\
                         & [606]11/2 & 0.195 & 24.3 & 0.18\\
\hline \\[-8pt]
$^{285}_{114}$Fl$_{171}$ & [604]9/2 & 0 & 22.8 & 0.17\\
                         & [611]1/2 & 0.009 & 22.2 & 0.17\\
                         & [611]3/2 & 0.139 & 21.5 & 0.16\\
                         & [606]11/2 & 0.169 & 22.2 & 0.17\\
                         & [613]5/2 & 0.184 & 21.4 & 0.16\\
                         & \textbf{[707]15/2} & 0.723 & 20.4 & 0.15\\
\hline \\[-8pt]
$^{287}_{116}$Lv$_{171}$ & [604]9/2 & 0 & 21.8 & 0.16\\
                         & [611]1/2 & 0.003 & 21.2 & 0.15\\
                         & [611]3/2 & 0.113 & 20.5 & 0.15\\
                         & [613]5/2 & 0.175 & 20.4 & 0.15\\
                         & [606]11/2 & 0.204 & 21.2 & 0.15\\
                         & \textbf{[707]15/2} & 0.661 & 19.2 & 0.14\\
\hline \\[-8pt]
$^{289}_{116}$Lv$_{173}$ & [611]1/2 & 0 & 18.1 & 0.13\\
                         & [611]3/2 & 0.256 & 17.7 & 0.13\\
                         & [613]5/2 & 0.394 & 17.8 & 0.13\\
                         & [604]9/2 & 0.649 & 18.4 & 0.13\\
                         & \textbf{[707]15/2} & 0.796 & 18.0 & 0.13\\
\end{tabular}
\end{ruledtabular}
\end{table}
\noindent

\subsection{Odd-odd nuclei}

By combining the low-lying 1-q.p.~excitations, one can deduce possible 2-q.p.~states in
the odd-odd nuclei that form $\alpha$-decay chains of $^{288}115$. It is worth
noting that there exist detailed calculations of 1-q.p. excitations in the
heaviest elements using the macroscopic-microscopic Woods-Saxon model
\cite{(Cwi94),(Par04),(Par05)}; unfortunately, they cannot be used to assess
the neutron s.p. structure in the region of interest as the range of neutron
numbers covered ($N\le 161$) in these papers is too limited.

\begin{table}[!htb]
\caption{\label{tabZ} Similar as in Table~\ref{tabN} but for
one-quasi-proton  excitations  in the $\alpha$-decay chains of $^{293}$117
and $^{287}$115 predicted with  {\UNEDFONESO}.
}
\begin{ruledtabular}
\begin{tabular}{clccc}
Nucleus & Config.  &  $E_x$ (MeV) & $Q_{20}$\,(b) & $\beta_2$ \\
\hline  \\[-8pt]
$^{279}_{111}$Rg$_{168}$ & [512]3/2 & 0 & 28.3 & 0.21\\
                         & [510]1/2 &  0.122 & 28.3 & 0.21\\
                         & \textbf{[615]11/2} & 0.284 & 28.4 & 0.21\\
                         & [505]9/2 & 0.392 & 27.7 & 0.20\\
                         & [521]1/2 & 0.633 & 26.6 & 0.19\\
\hline  \\[-8pt]
$^{281}_{111}$Rg$_{170}$ & [512]3/2 & 0 & 26.0 & 0.20\\
                         & [510]1/2 & 0.165 & 26.1 & 0.20\\
                         & [505]9/2 & 0.263 & 25.8 & 0.20\\
                         & \textbf{[615]11/2} & 0.277 & 26.1 & 0.20\\
                         & [521]1/2 & 0.412 & 25.3 & 0.19\\
\hline  \\[-8pt]
$^{283}$113$_{170}$      & [512]3/2 & 0 & 24.5 & 0.18\\
                         & [510]1/2 & 0.057 & 24.5 & 0.18\\
                         & [505]9/2 & 0.090 & 24.4 & 0.18\\
                         & [503]7/2 & 0.453 & 22.9 & 0.16\\
                         & [521]1/2 & 0.629 & 23.2 & 0.16\\
\hline  \\[-8pt]
$^{285}$113$_{172}$      & [512]3/2 & 0 & 18.7 & 0.14\\
                         & [503]7/2 & 0.096 & 17.6 & 0.13\\
                         & [550]1/2 & 0.146 & 17.2 & 0.13\\
                         & [510]1/2 & 0.177 & 19.4 & 0.15\\
                         & [505]9/2 & 0.380 & 19.7 & 0.15\\
                         & \textbf{[615]11/2} & 0.645 & 18.2 & 0.14\\
\hline  \\[-8pt]
$^{287}115_{172}$   & [512]3/2 & 0 & 14.4 & 0.10\\
                         & [550]1/2 & 0.042 & 12.6 & 0.09\\
                         & [503]7/2 & 0.127 & 14.2 & 0.10\\
                         & \textbf{[606]13/2} & 0.206 & 14.2 & 0.10\\
                         & [510]1/2 & 0.388 & 14.1 & 0.10\\
\hline  \\[-8pt]
$^{289}$115$_{174}$ & [550]1/2 & 0 & 11.3 & 0.08\\
                         & [512]3/2 & 0.002 & 11.5 & 0.08\\
                         & \textbf{[606]13/2} & 0.137 & 12.2 & 0.09\\
                         & [503]7/2 & 0.230 & 11.6 & 0.08\\
                         & [510]1/2 & 0.338 & 11.6 & 0.08\\
\hline  \\[-8pt]
$^{293}$117$_{176}$ & [512]3/2 & 0 & 7.8 & 0.05\\
                         & [550]1/2 & 0.06 & 7.5 & 0.05\\
                         & [510]1/2 & 0.218 & 7.8 & 0.05\\
                         & [503]5/2 & 0.310 & 7.0 & 0.04\\
                         & [503]7/2 & 0.778 & 8.1 & 0.06
\end{tabular}
\end{ruledtabular}
\end{table}
\begin{table}[!htb]
\caption{\label{tabZ-ori} Similar as in Table~\ref{tabZ} but with {\UNEDFONE}.
}
\begin{ruledtabular}
\begin{tabular}{clccc}
Nucleus & Config.  &  $E_x$ (MeV) & $Q_{20}$\,(b) & $\beta_2$ \\
\hline \\[-8pt]
$^{279}_{111}$Rg$_{168}$ & [512]3/2 & 0 & 27.8 & 0.21\\
                         & \textbf{[615]11/2} &  0.075 & 27.8 & 0.21\\
                         & [505]9/2 & 0.119 & 27.6 & 0.21\\
                         & [510]1/2 & 0.252 & 27.8 & 0.21\\
                         & \textbf{[624]9/2} & 0.793 & 27.6 & 0.21\\
\hline \\[-8pt]
$^{281}_{111}$Rg$_{170}$ & \textbf{[615]11/2} & 0 & 25.3 & 0.19\\
                         & [512]3/2 & 0.021 & 25.3 & 0.19\\
                         & [505]9/2 & 0.087 & 25.4 & 0.19\\
                         & [510]1/2 & 0.218 & 25.3 & 0.19\\
                         & [521]1/2 & 0.601 & 23.9 & 0.18\\
\hline \\[-8pt]
$^{283}$113$_{170}$      & [510]1/2 & 0 & 24.3 & 0.18\\
                         & [512]3/2 & 0.024 & 24.1 & 0.18\\
                         & \textbf{[615]11/2} & 0.055 & 24.3 & 0.18\\
                         & [503]7/2 & 0.480 & 22.7 & 0.17\\
                         & [550]1/2 & 0.702 & 22.7 & 0.17\\
\hline \\[-8pt]
$^{285}$113$_{172}$      & [510]1/2 & 0 & 21.2 & 0.16\\
                         & [512]3/2 & 0.044 & 21.0 & 0.15\\
                         & \textbf{[615]11/2} & 0.089 & 21.1 & 0.15\\
                         & [505]9/2 & 0.287 & 22.8 & 0.16\\
                         & [550]1/2 & 0.411 & 19.8 & 0.15\\
\hline \\[-8pt]
$^{287}$115$_{172}$      & [510]1/2 & 0 & 20.2 & 0.15\\
                         & [503]7/2 & 0.055 & 20.1 & 0.15\\
                         & [512]3/2 & 0.193 & 20.1 & 0.15\\
                         & \textbf{[615]11/2} & 0.253 & 20.4 & 0.15\\
                         & [550]1/2 & 0.682 & 19.1 & 0.14\\
\hline \\[-8pt]
$^{289}$115$_{174}$      & [503]7/2 & 0 & 17.1 & 0.12\\
                         & [512]3/2 & 0.024 & 16.9 & 0.12\\
                         & [510]1/2 & 0.085 & 17.1 & 0.12\\
                         & \textbf{[615]11/2} & 0.272 & 17.3 & 0.12\\
                         & [550]1/2 & 0.438 & 16.5 & 0.12\\
\hline \\[-8pt]
$^{293}$117$_{176}$      & [510]1/2 & 0 & 11.4 & 0.08\\
                         & \textbf{[606]13/2} & 0.003 & 11.5 & 0.08\\
                         & [512]3/2 & 0.03 & 11.2 & 0.08\\
                         & [550]1/2 & 0.197 & 9.7 & 0.06\\
                         & [503]5/2 & 0.26 & 11.1 & 0.08\\
                         & [503]7/2 & 0.266 & 11.7 & 0.08\\
                         & \textbf{[615]11/2} & 0.588 & 11.4 & 0.08\\
\end{tabular}
\end{ruledtabular}
\end{table}

\begin{figure}[htb]
\includegraphics[width=0.7\columnwidth]{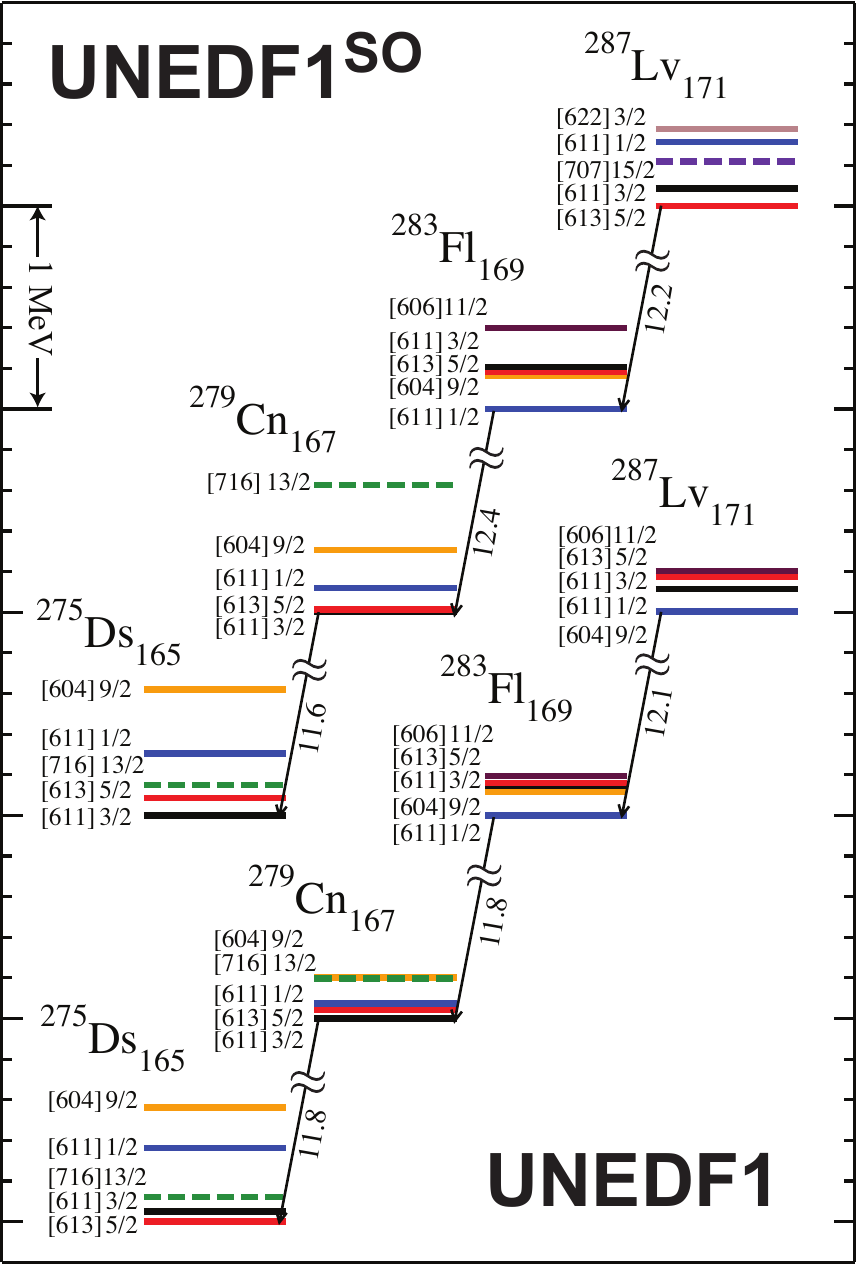}
\caption{\label{figN} (Color online)
1-q.p.~spectra for nuclei forming the $\alpha$-decay chain $^{287}$Lv
$\rightarrow \cdots \rightarrow$ $^{275}$Ds predicted with {\UNEDFONESO} (upper
sequence) and {\UNEDFONE} (lower sequence).  $Q_{\alpha}$ values for
g.s.$\rightarrow$g.s. transitions are marked. The binding energy differences
between different nuclei are shifted arbitrarily, whereas the excitation
energies within a given nucleus are shown to scale.
} \end{figure}

\begin{figure}[htb]
\includegraphics[width=0.7\columnwidth]{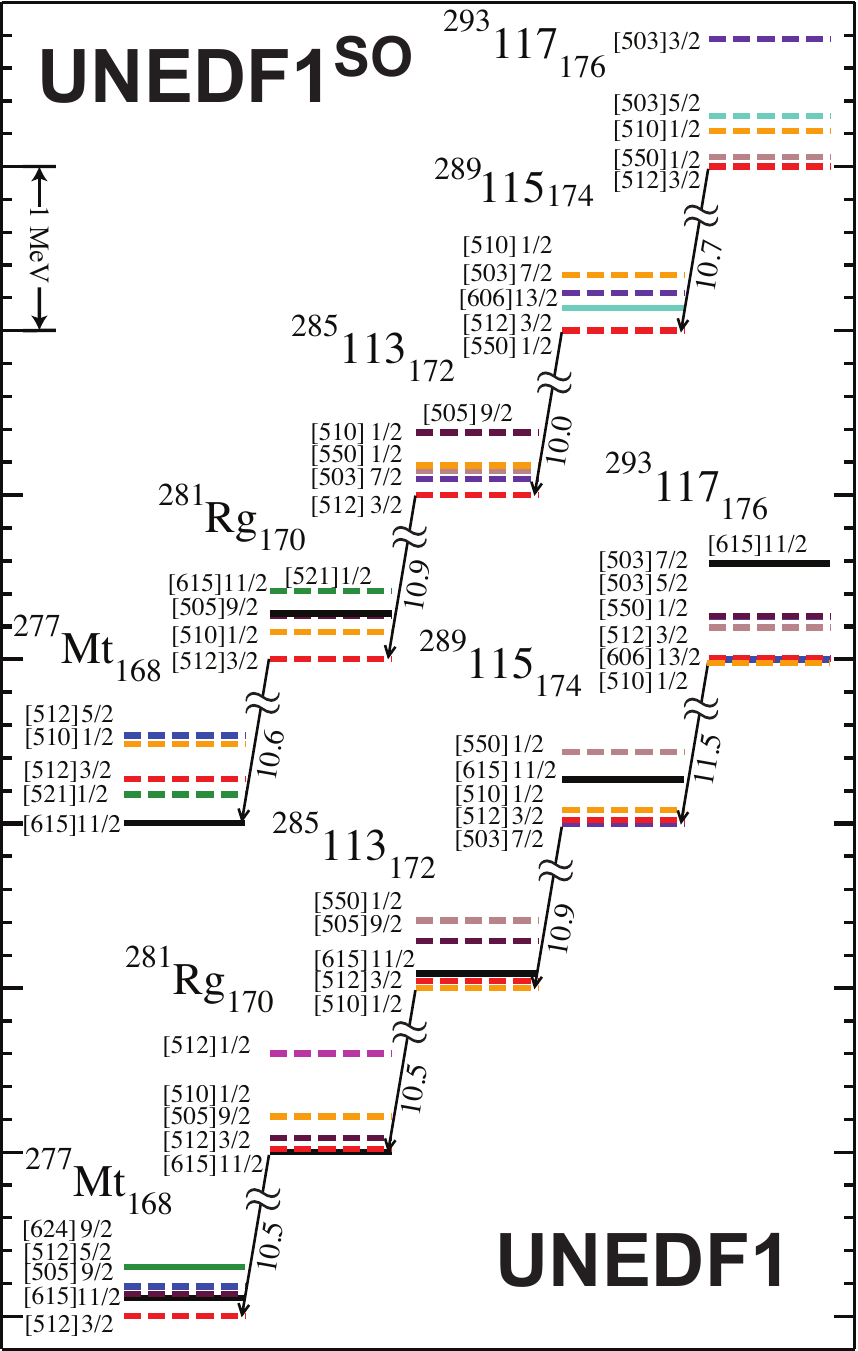}
\caption{\label{figZ} (Color online)
Similar as  in Fig.~\ref{figN} but for the $\alpha$-decay chain $^{293}117$ $\rightarrow \cdots \rightarrow$ $^{277}$Mt.
}
\end{figure}

\begin{table}[htb]
\caption{\label{tabNZMT} Similar as in Table~\ref{tabN} but for
one-quasi-particle  excitations  in the
odd-$A$ neighbors of $^{276}_{109}$Mt$_{167}$
predicted with {\UNEDFONESO} and {\UNEDFONE}.
}
\begin{ruledtabular}
\begin{tabular}{clccc}
Nucleus & Config.  &  $E_x$ (MeV) & $Q_{20}$\,(b) & $\beta_2$ \\
\hline  \\[-8pt]
&&{\UNEDFONESO}&& \\
$^{275}_{109}$Mt$_{166}$ & \textbf{[615]11/2} & 0 & 29.7 & 0.23\\
                       & [512]3/2 & 0.243 & 29.5 & 0.23 \\
                       & [521]1/2 & 0.402 & 29.2 & 0.22 \\
                       & [512]5/2 & 0.500 & 29.7 & 0.23\\
                       & [510]1/2 & 0.512 & 29.7 & 0.23\\
$^{277}_{109}$Mt$_{168}$ & \textbf{[615]11/2} & 0 & 28.3 & 0.23\\
                      & [521]1/2 & 0.174 & 27.4 & 0.21\\
                         & [512]3/2 & 0.269 & 28.3 & 0.23\\
                         & [510]1/2 & 0.480 & 28.3 & 0.23\\
                         & [512]5/2 & 0.532 & 28.3 & 0.23\\
$^{275}_{108}$Hs$_{167}$ & [613]5/2 & 0 & 28.9 & 0.22\\
                          & [611]3/2 & 0.016 & 28.9 & 0.22\\
                         & [611]1/2 & 0.117 & 28.9 & 0.22\\
                         & [604]9/2 & 0.235 & 28.2 & 0.22 \\
                         & \textbf{[716]13/2} & 0.479 & 29.4 & 0.23 \\
$^{277}_{110}$Ds$_{167}$ & [611]3/2 & 0 & 28.6 & 0.22\\
                         & [613]5/2 & 0.046 & 28.9 & 0.22\\
                         & [611]1/2 & 0.107 & 28.7 & 0.22\\
                         & [604]9/2 & 0.335 & 28.6 & 0.22\\
                         & \textbf{[716]13/2} & 0.564 & 29.2 & 0.22\\
\hline  \\[-8pt]
&&{\UNEDFONE}&& \\
$^{275}_{109}$Mt$_{166}$ & [512]3/2 & 0 & 30.0 & 0.23\\
                         & \textbf{[615]11/2} & 0.159 & 29.6 & 0.23\\
                         & [505]9/2 & 0.167 & 29.0 & 0.22 \\
                         & [510]1/2 & 0.173 & 30.1 & 0.23\\
                         & \textbf{[624]9/2} & 0.318 & 29.9 & 0.23 \\
$^{277}_{109}$Mt$_{168}$ & [512]3/2 & 0 & 28.6 & 0.22\\
                         & \textbf{[615]11/2} & 0.131 & 28.2 & 0.21\\
                         & [505]9/2 & 0.136 & 27.5 & 0.21\\
                         & [512]5/2 & 0.182 & 28.6 & 0.22\\
                         & \textbf{[624]9/2} & 0.300 & 28.5 & 0.21\\
$^{275}_{108}$Hs$_{167}$ & [613]5/2 & 0 & 29.4 & 0.22\\
                         & [611]3/2 & 0.067 & 29.4 & 0.22\\
                         & [611]1/2 & 0.104 & 29.5 & 0.22\\
                         & \textbf{[716]13/2} & 0.173 & 29.7 & 0.23 \\
                         & [604]9/2 & 0.242 & 29.1 & 0.22 \\
$^{277}_{110}$Ds$_{167}$ & [611]3/2 & 0 & 28.9 & 0.22\\
                         & [613]5/2 & 0.035 & 29.0 & 0.22\\
                         & [611]1/2 & 0.090 & 29.0 & 0.22\\
                         & \textbf{[716]13/2} & 0.157 & 29.2 & 0.22\\
                         & [604]9/2 & 0.227 & 28.7 & 0.22\\
\end{tabular}
\end{ruledtabular}
\end{table}
Let us look into the structure of $^{276}$Mt in some detail. The structural
information relevant to this nucleus is contained in the 1-q.p.~spectra of its
odd-$A$ neighbors $^{275,277}$Mt, $^{275}$Hs, and $^{277}$Ds, provided in SEDF
Table~\ref{tabNZMT}. All low-lying 1-q.p.~states in these nuclei
correspond to very similar quadrupole mass deformation of $\beta_2\approx
0.22$, which facilitates comparison with the Nilsson diagram of
Fig.~\ref{figEspEDF}. The lowest 1-q.p.~proton states are the unique-parity
[615]11/2 and $\Nosc=5$ excitations [512]3/2, [521]1/2, [510]1/2, and [512]5/2.
The 1-q.p.~neutron structure corresponds to the [716]13/2 intruder and
$\Nosc=6$ [611]3/2, [613]5/2, [611]1/2, and [604]9/2 Nilsson orbits. The most
significant difference between the two SEDF models is the appearance of the
[505]9/2 1-q.p.~proton excitation low in energy in {\UNEDFONE}. According to
the MO model of Fig.~\ref{figEspNS}, the lowest 1-q.p proton excitations are
the [615]11/2, [521]1/2, and [505]9/2 Nilsson orbits, while the lowest neutron
states are: [606]11/2, [604]9/2, [611]3/2, [613]5/2, and [716]13/2. It is
interesting to note that the structure of 1-q.p. proton states predicted for
$^{275}$Mt in Ref.~\cite{(Par04)} falls between predictions of {\UNEDFONE} and
{\UNEDFONESO}. In addition, within the Woods-Saxon model of
Ref.~\cite{(Cha77)}, the proton states [615]11/2 and [505]9/2 are
the two lowest orbitals for Mt over a large range of deformations.

\begin{figure}[htb]
\includegraphics[width=\columnwidth]{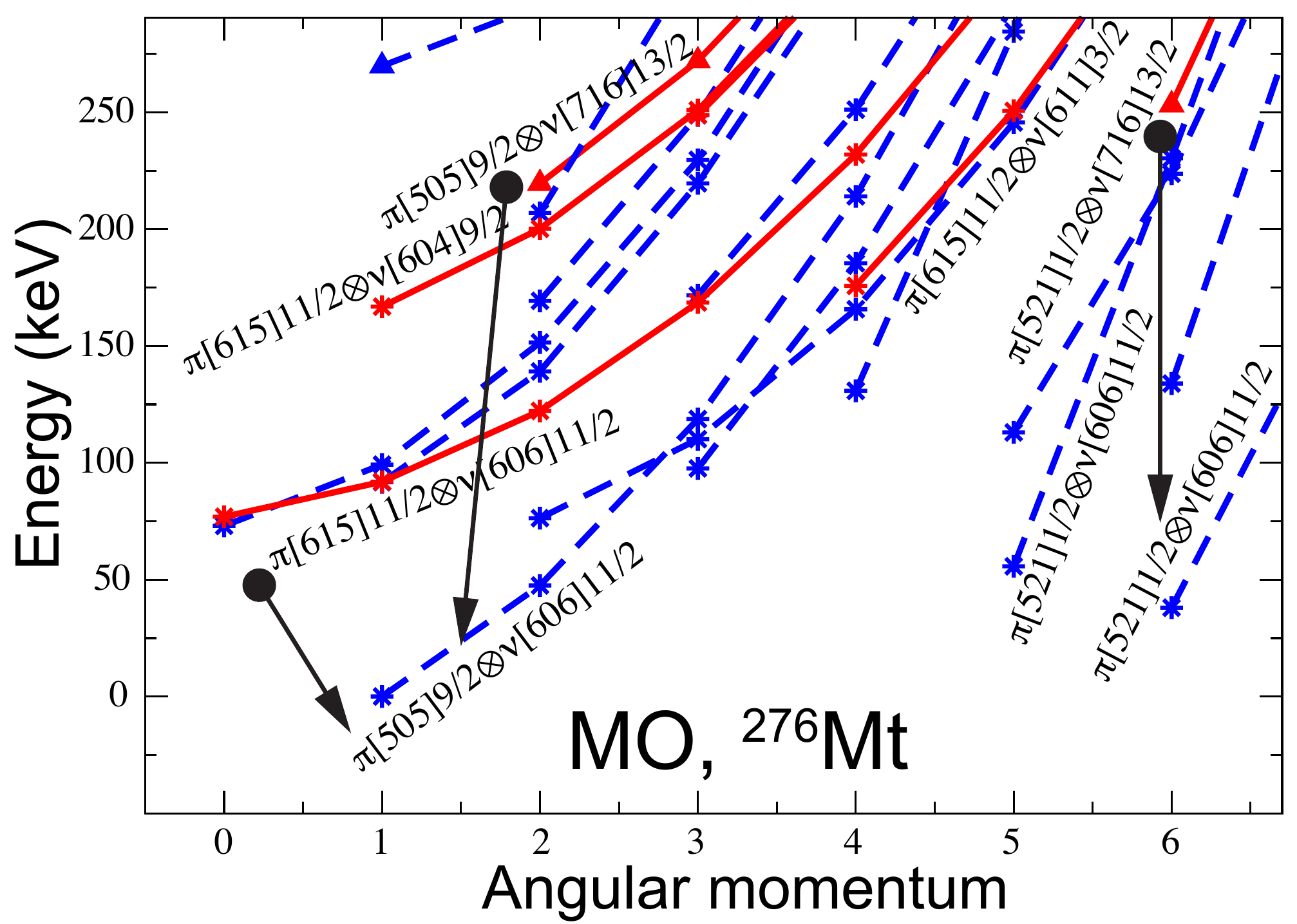}
\caption{\label{p+rotMt} (Color online)
Results of 2-q.p.-plus-rotor NS calculations for $^{276}$Mt. States connected
with lines have the same dominating single-particle configurations; they can be
interpreted as rotational bands built on 2-q.p. bandheads indicated.
Solid/dashed lines mark bands with positive/negative parity. Positive/negative
parity neutron configurations are indicated by asterisks/triangles. The
candidates for stretched $\Delta \Omega=0,\pm 1$ $E1$ transitions are marked by
thick arrows.
}
\end{figure}

As noted in \cite{(Rud13a)}, there is a very limited choice of q.p.
configurations that could generate the observed $E1$ transitions in $^{276}$Mt.
If one insists on a strict conservation of the  $\Omega$ quantum number for
protons and neutrons, no low-energy $E1$ transitions are predicted by
{\UNEDFONESO}. Formally, one can construct states that can be connected by an
$\Delta \Omega=0,\pm 1$, parity changing operator, e.g.,
$\{\pi[615]11/2\otimes\nu[613]5/2\}_{3^+}$ and
$\{\pi[521]1/2\otimes\nu[613]5/2\}_{2^-,3^-}$, but a significant Coriolis
coupling would be required to produce a measurable $E1$ rate. The situation is
fairly straightforward with {\UNEDFONE}. Here, the stretched $E1$ transition
$\pi[505]9/2 \rightarrow \pi[615]11/2$ can explain the data, with the neutron
spectator orbital being [611]3/2 or [613]5/2 or [611]1/2. The NS approach
predicts two scenarios: the proton $\pi[615]11/2 \rightarrow \pi[505]9/2$
transition as in {\UNEDFONE} and the neutron $\nu[716]13/2 \rightarrow
\nu[606]11/2$ transition. According to 2-q.p.-plus-rotor calculations shown in
Fig.~\ref{p+rotMt} both scenarios are equally likely. It is 
interesting to note  that the splitting between the
$I=1$ and $I=2$ members of the lowest $K^{\pi}=1^-$
($\pi[505]9/2\otimes\nu[606]11/2$) band is only 43\,keV, see  Fig.~\ref{p+rotMt}. This is particularly close to the energy difference $\approx47$\,keV 
between the states suggested experimentally \cite{(Rud13a)}. 

\begin{figure}[htb]
\includegraphics[width=\columnwidth]{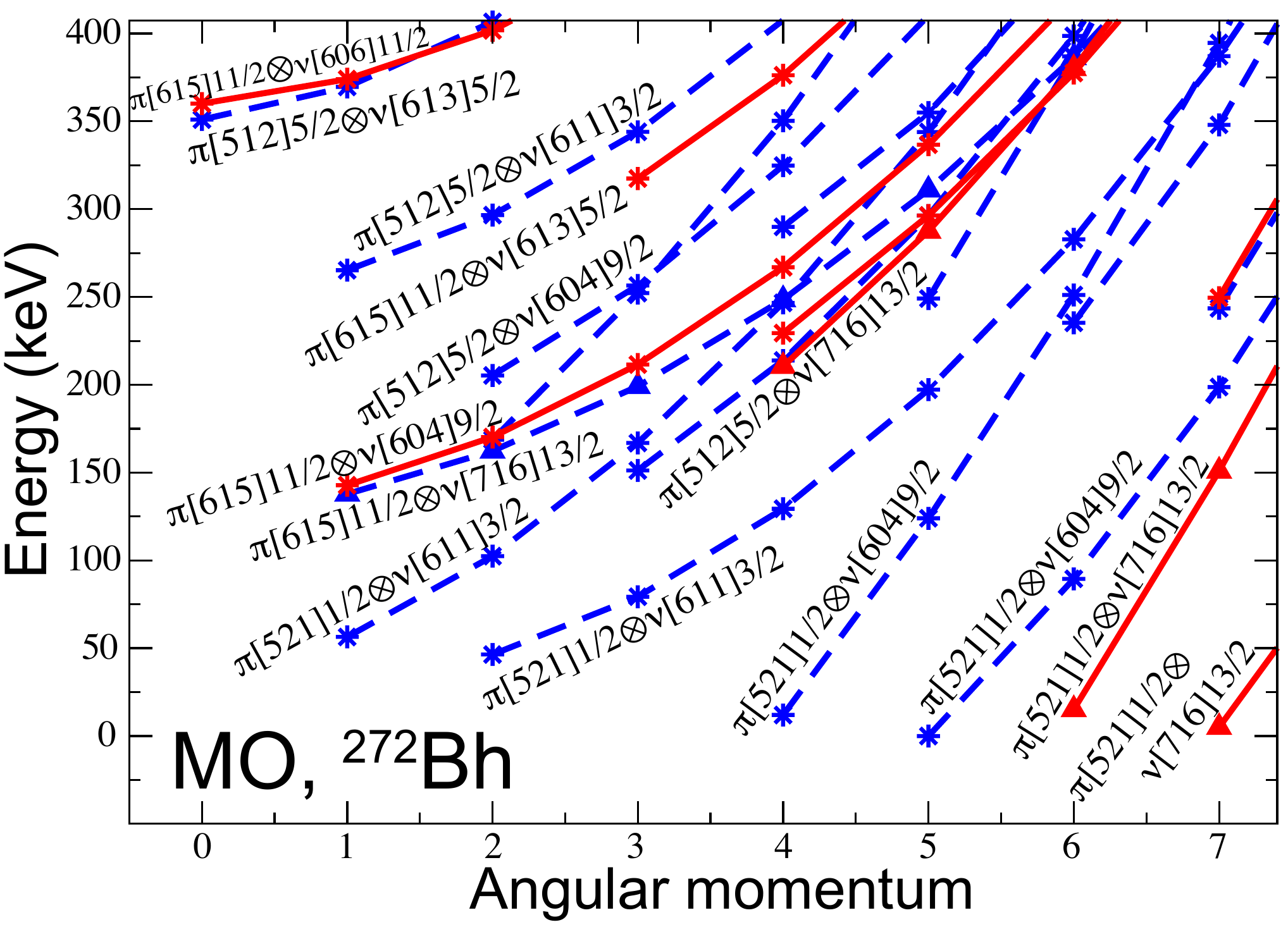}
\caption{\label{p+rotBh} (Color online)
Similar as in Fig.~\ref{p+rotMt} but for $^{272}$Bh.
}
\end{figure}

To analyse the case of $^{272}$Bh, we have calculated the 1-q.p.~spectra of
$^{271,273}$Bh, $^{273}$Hs, and $^{271}$Sg, see Table~\ref{tabNZBH}. The resulting level scheme is very
complex. Indeed, as seen in in Fig.~\ref{p+rotBh}, 2-q.p.-plus-rotor
calculations predict quite a few candidates for low-energy $E2$ and $M1$
transitions, and this is consistent with experiment \cite{(Rud13a)}. The
results displayed in Figs.~\ref{p+rotMt} and ~\ref{p+rotBh} clearly demonstrate
that a small change in ordering of Nilsson orbits can influence the decay
scenarios.
\begin{table}[htb]
\caption{\label{tabNZBH} Similar as in Table~\ref{tabNZMT} but for
one-quasi-particle  excitations  in the
odd-$A$ neighbors of $^{272}_{107}$Bh$_{165}$
predicted with {\UNEDFONESO} and {\UNEDFONE}.
}
\begin{ruledtabular}
\begin{tabular}{clccc}
Nucleus & Config.  &  $E_x$ (MeV) & $Q_{20}$\,(b) & $\beta_2$ \\
\hline  \\[-8pt]
&&{\UNEDFONESO}&& \\
$^{271}_{107}$Bh$_{164}$ & [512]5/2 & 0 & 30.8 & 0.24\\
                   & [521]1/2 & 0.076 & 30.3 & 0.23\\
                   & \textbf{[615]11/2} & 0.244 & 30.4 & 0.23\\
                   & \textbf{[624]9/2} & 0.415 & 30.8 & 0.24\\
                   & [514]7/2 & 0.682 & 31.1 & 0.24\\
$^{273}_{107}$Bh$_{166}$       & \textbf{[615]11/2} & 0 & 29.3 & 0.23\\
                         & [512]5/2 & 0.002 & 29.6 & 0.23\\
                         & [521]1/2 & 0.072 & 29.5 & 0.23\\
                         & \textbf{[624]9/2} & 0.243 & 29.6 & 0.23\\
                         & [512]3/2 & 0.574 & 29.4 & 0.23\\
                         & [514]7/2 & 0.692 & 29.8 & 0.23\\
$^{271}_{106}$Sg$_{165}$ & [611]3/2 & 0 & 30.1 & 0.23\\
                   & \textbf{[716]13/2} & 0.156 & 30.2 & 0.23\\
                   & [613]5/2 & 0.223 & 30.1 & 0.23\\
                   & [611]1/2 & 0.311 & 30.1 & 0.23\\
                   & [604]9/2 & 0.429 & 29.2 & 0.23\\
$^{273}_{108}$Hs$_{165}$ & [611]3/2 & 0 & 30.1 & 0.23\\
                   & [613]5/2 & 0.118 & 30.0 & 0.23\\
                   & \textbf{[716]13/2} & 0.142 & 30.3 & 0.23\\
                   & [611]1/2 & 0.314 & 30.1 & 0.23\\
                   & [604]9/2 & 0.518 & 29.2 & 0.22\\
\hline \\[-8pt]
&&{\UNEDFONE}&& \\
$^{271}_{107}$Bh$_{164}$ & [512]5/2 & 0 & 31.6 & 0.24\\
                   & \textbf{[624]9/2} & 0.028 & 31.5 & 0.24\\
                   & [512]3/2 & 0.338 & 31.3 & 0.24\\
                   & [521]1/2 & 0.554 & 30.9 & 0.24\\
                   & [505]9/2 & 0.714 & 30.0 & 0.23\\
$^{273}_{107}$Bh$_{166}$       & \textbf{[624]9/2} & 0 & 30.7 & 0.23\\
                       & [512]5/2 & 0.076 & 30.7 & 0.23\\
                         & [512]3/2 & 0.306 & 30.6 & 0.23\\
                         & [521]1/2 & 0.377 & 29.6 & 0.23\\
                         & [505]9/2 & 0.599 & 28.7 & 0.22\\
$^{271}_{106}$Sg$_{165}$ & [611]3/2 & 0 & 31.1 & 0.24\\
                   & [613]5/2 & 0.007 & 30.7 & 0.24\\
                   & \textbf{[716]13/2} & 0.11 & 31.1 & 0.24\\
                   & [611]1/2 & 0.252 & 31.0 & 0.24\\
                   & [604]9/2 & 0.635 & 33.2 & 0.25\\
$^{273}_{108}$Hs$_{165}$ & [613]5/2 & 0 & 30.7 & 0.24\\
                   & [611]3/2 & 0.048 & 30.8 & 0.24\\
                   & \textbf{[716]13/2} & 0.119 & 30.8 & 0.24\\
                   & [611]1/2 & 0.300 & 30.8 & 0.24\\
                   & [604]9/2 & 0.508 & 30.4 & 0.23\\
\end{tabular}
\end{ruledtabular}
\end{table}

The calculated $Q_{\alpha}$ values depend, of course, on the structure of
parent and daughter states \cite{(Cwi99)} (see Figs.~\ref{figN} and
\ref{figZ}). The agreement with the measured values for the heaviest elements
is reasonable, usually better than 1\,MeV. This is comparable with other
calculations \cite{(Ben13),(Pra12),(War12),(Sta13)}.

\section{Conclusions}

In summary, we studied shell structure of superheavy nuclei within the
self-consistent SHFB approach and macroscopic-microscopic NS model.  Detailed
predictions have been made for the quasi-proton and quasi-neutron structures of
nuclei belonging to the $\alpha$-decay chains of $^{287}115$, $^{287}$Lv,
$^{289}$Lv, and $^{293}$117. The {\UNEDFONE} and {\UNEDFONESO} SEDF models
differ in the strength of the spin-orbit term, and this impacts detailed
predictions for the deformed nuclei around $Z=110$ and $N=168$. The recent
observation of low-energy $E1$ transitions in $^{276}$Mt \cite{(Rud13a)}
provides a stringent constraint on theoretical models. Indeed, the recently
proposed {\UNEDFONESO} parametrization that performs well in the transfermium
region does not offer a simple explanation of the $E1$ data, whereas the global
{\UNEDFONE} parametrization explains the data in terms of the proton
$\pi[505]9/2 \rightarrow \pi[615]11/2$ transition. The MO models suggests two
competing scenarios: a proton transition similar to that of {\UNEDFONE}, and an
alternative neutron $\nu [716]13/2 \rightarrow \nu [606]11/2$ $E1$ transition.
To confirm or disprove these scenarios, theory strongly recommends a search
for $E1$ transitions in neighboring odd-$A$ systems $^{275,277}$Mt, $^{275}$Hs,
and $^{277}$Ds. Experimentally, this calls for high-resolution $\alpha$-photon
coincidence spectroscopy of decay chains starting from $^{293}$117,
$^{287,289}$115, or $^{285,287}$Fl, respectively. However, the observation of
these systems is hampered either by relatively low production cross-sections or
large spontenous fission branches on the way to the nuclei of structural
interest
\cite{(Oga07),(Oga11),(Oga12),(Oga12a),(Oga13),(Dul10),(Gat11),(Hof12),(Rud13a)}.
A solution to this spectroscopic puzzle may significantly contribute to our
understanding of shell structure in superheavy nuclei, and the strength of the
spin-orbit splitting in particular.

\bigskip
\begin{acknowledgments}
Discussions with S. {\AA}berg are
gratefully acknowledged.
This work was supported by the U.S. Department of
Energy (DOE) under Contracts No.
DE-FG02-96ER40963
(University of Tennessee),  No.
DE-SC0008499    (NUCLEI SciDAC Collaboration),
 No. DE-NA0001820 (the Stewardship
Science Academic Alliances program);
by the Academy of Finland and University of
Jyv\"askyl\"a within the FIDIPRO programme;
 by the
Polish National Science Center under Contract No.~2012/07/B/ST2/03907; and by the Swedish Research Council. An award of computer
time was provided by the National Institute for Computational
Sciences (NICS) and the Innovative and Novel Computational Impact on
Theory and Experiment (INCITE) program using resources of the OLCF facility.
\end{acknowledgments}

\bibliographystyle{apsrev4-1}

%

\end{document}